\documentclass[aps,a4paper,floatfix, nofootinbib]{revtex4}

\usepackage[utf8]{inputenc}

\usepackage{graphicx,color}
\usepackage{amsmath,amssymb,amsfonts,amsthm,amscd,bm}
\usepackage{booktabs}
\usepackage{cancel}

\def\bar{\overline}
\def\hat{\widehat}
\def\*{\star}
\def\[{\left[}
\def\]{\right]}
\def\({\left(}      

\def\){\right)}

\def\zbar{{\bar{z} }}
\def\frac#1#2{\dfrac{#1}{#2}}
\def\inv#1{\dfrac{1}{#1}}

\def\d{\partial}

\def\2pi{\hbox{$2\pi i$}}

\def\dsl{\raise.15ex\hbox{/}\kern-.57em\partial}
\def\Dsl{\,\raise.15ex\hbox{/}\mkern-.13.5mu D}
\def\vep{\varepsilon}

\def\CA{{\cal A}}      
   \def\CE{{\cal E}}   \def\CF{{\cal F}}

   \def\CN{{\cal N}}   
      
\def\CS{{\cal S}}      \def\CU{{\cal U}}

\def\2pi{\hbox{$2\pi i$}}

\def\dsl{\raise.15ex\hbox{/}\kern-.57em\partial}
\def\Dsl{\,\raise.15ex\hbox{/}\mkern-.13.5mu D}
\font\numbers=cmss12
\font\upright=cmu10 scaled\magstep1
\def\stroke{\vrule height8pt width0.4pt depth-0.1pt}
\def\topfleck{\vrule height8pt width0.5pt depth-5.9pt}
\def\botfleck{\vrule height2pt width0.5pt depth0.1pt}
\def\Zmath{\vcenter{\hbox{\numbers\rlap{\rlap{Z}\kern
    0.8pt\topfleck}\kern 2.2pt
    \rlap Z\kern 6pt\botfleck\kern 1pt}}}
\def\Qmath{
    \vcenter{\hbox{\upright\rlap{\rlap{Q}\kern3.8pt\stroke}\phantom{Q}}}}
\def\Nmath{\vcenter{\hbox{\upright\rlap{I}\kern 1.7pt N}}}
\def\Cmath{\vcenter{\hbox{\upright\rlap{\rlap{C}\kern
                   3.8pt\stroke}\phantom{C}}}}
\def\Rmath{\vcenter{\hbox{\upright\rlap{I}\kern 1.7pt R}}}
\def\Z{\ifmmode\Zmath\else$\Zmath$\fi}
\def\Q{\ifmmode\Qmath\else$\Qmath$\fi}
\def\N{\ifmmode\Nmath\else$\Nmath$\fi}
\def\C{\ifmmode\Cmath\else$\Cmath$\fi}
\def\R{\ifmmode\Rmath\else$\Rmath$\fi}
\def\barray{\begin{eqnarray}}
\def\earray{\end{eqnarray}}
\def\beq{\begin{equation}}
\def\eeq{\end{equation}}

\def\kvec{{\bf{k}}}

\def\smallhalf{{\scriptstyle \inv{2}}}

\def\smallhalf{{\textstyle \inv{2}}}

\def\AA{\leavevmode\setbox0=\hbox{h}
\dimen0=\ht0 \advance\dimen0 by-1ex\rlap{\raise.67\dimen0\hbox{\char'27}}A}

\makeatletter
\def\iddots{\mathinner{\mkern1mu\raise\p@
\vbox{\kern7\p@\hbox{.}}\mkern2mu
\raise4\p@\hbox{.}\mkern2mu\raise7\p@\hbox{.}\mkern1mu}}
\makeatother

\def\Tbar{\bar{T}}

\def\vep{\varepsilon}

\theoremstyle{plain}

\theoremstyle{remark}

\def\Tvac{\langle T_{\mu \nu} \rangle_0}

\begin{document}

\def\cstarUV{c^*_{UV}}

\def\kvec{{\bf k}}

\def\rhovac{\rho_{\rm vac}}
\def\smallhalf{\tfrac{1}{2}}

\def\gcal{\mathfrak{g} }

\def\cIR{c_{\rm IR}}
\def\cUV{c_{\rm UV}}

\def\cftIR{{\rm cft}_{\rm IR}}
\def\cftUV{ {\rm cft}_{\rm UV}}

\def\cbulk{c_{\rm bulk}}

\def\ceff{c_{\rm eff}}

\def\cpert{c_{\rm pert}}

\title{
Mingling  of the infrared and ultraviolet and the ``cosmological constant"  for interacting QFT in 2d
}
\author{
 Andr\'e  LeClair\footnote{andre.leclair@gmail.com}
}
\affiliation{Cornell University, Physics Department, Ithaca, NY 14853, United States of America} 

\begin{abstract}

We propose  a proper definition of the vacuum expectation value of the stress energy tensor $\langle 0 | T_{\mu\nu} |0 \rangle$ for integrable quantum field theories in two spacetime dimensions,   which is the analog of the cosmological constant in 4d.     For a wide variety of models,   massive or massless,    we show $\rhovac = - m^2/2\gcal$ exactly,     where $\gcal$ is a generalized coupling which we compute 
 and $m$ is 
a basic mass scale.    The kinds of models we consider are  the  massive  sinh-Gordon and sine-Gordon theories and perturbations of the
Yang-Lee and 3-state Potts models,   pure $T\Tbar$ perturbations of infra-red QFT's,   and UV completions of the latter which are massless flows between UV and IR fixed points.   
 In the massive case $m$ is the  physical mass of the {\it  lightest}  particle and $\gcal$ is related to parameters in the  2-body 
S-matrix.  In some examples $\rhovac =0$ due to a fractional supersymmetry.
  For massless cases,  $m$ can be a scale of spontaneous symmetry breaking.    
The ``cosmological constant problem"  generically  arises in the free field limit $\gcal \to 0$,   thus interactions can potentially resolve the problem at least for most cases considered in this paper.      
We speculate on extensions of these results to 4 spacetime dimensions and propose $\rhovac = m^4/2 \gcal$,
 however without integrability we cannot yet propose a precise manner in which to calculate $\gcal$.   Nevertheless,    based on cosmological data on $\rhovac$,  if $\gcal \sim 1$ then it is worth pointing out that  the lightest mass particle is on the order of 
experimental values of  proposed neutrino masses.

\end{abstract}

\maketitle
\tableofcontents

\section{Introduction}

This work concerns the proper definition of the vacuum expectation value of the stress-energy tensor $T_{\mu \nu}$ for integrable quantum field theory in two spacetime dimensions for flat Minkowski space.    The calculations presented shed some light on the so-called cosmological constant problem,  so in this Introduction let us motivate this  work based on this connection.     Einstein's equations of general relativity involve the classical stress-energy tensor as a source of gravitation.     It is a natural idea that in a semi-classical quantum theory  the classical $T_{\mu\nu}$  is replaced by its quantum vacuum expectation value $\Tvac = \langle 0 | T_{\mu\nu}  | 0 \rangle$, where $|0\rangle$ is the vacuum state.
Based on general coordinate invariance one expects
\beq
\label{CC1}
\Tvac = - \rhovac \, g_{\mu\nu}
\eeq
where $g_{\mu\nu}$ is the spacetime metric.    In the above equation the convention is that in flat space $g_{00} =  -g_{11} = -1$ such that
$\langle 0| T_{00}  |0 \rangle$ is an energy density.            
In its simplest form \cite{Weinberg},  the cosmological constant problem is that  the zero point energy of a {\it free} quantum field for a particle of mass $m$,   which is a collection of harmonic oscillators of frequency $\omega_k = \sqrt{\kvec^2 + m^2}$, is naively given by 
\beq
\label{CC2}
\rhovac =  \int_0^\Lambda \frac{dk}{(2 \pi )^3 } \,  4 \pi k^2   ~ \smallhalf  \sqrt{k^2  + m^2}  \approx   \frac{\Lambda^4}{16 \pi^2} - \frac{m^4}{32 \pi^2 } \log (2 \Lambda/m)  
\eeq
where $\Lambda$ is an ultraviolet cutoff and we have assumed $\Lambda\gg m$.     The problem is that for some reasonable choices of  $\Lambda$,  such as the Planck scale,  or perhaps of lower scales based on spontaneous symmetry breaking,   $\rhovac$ is huge.   Depending on definitions and comparison to data,   $\rhovac$ is off by perhaps over 100 orders of magnitude compared to experimental cosmological values.    Nature is very strongly suggesting that  this is likely  a  major conceptual problem.     Indeed there are many issues with the above calculation. 
  First is that perhaps the zero point energy is not a source of gravitation at all,  since in both classical and quantum physics without gravity   the zero point energy is not absolutely defined. 
   It is also possible that the observed small cosmological constant does not come from  $\langle 0 | T_{\mu\nu} |0\rangle$ in the first place,  but rather from something not understood in classical gravity.   
We take the conventional point of view that $\langle T_{\mu\nu} \rangle_0 $  can be well  defined,  and  that all contributions to $T_{\mu \nu}$ gravitate,    and thus focus on a proper physical definition of $\rhovac$.     In quantum field theory (QFT) one is accustomed to dealing with ultraviolet (UV) divergences  as $\Lambda \to \infty$ with  various regularization and renormalization methods,  which suggests that the computation in \eqref{CC2} is not physically meaningful by itself,   if at all,   and this is the attitude  that we take here.

 Henceforth,  we depart from cosmology per se,  and focus on defining $\rhovac$ for integrable QFT's in 2 spacetime dimensions without gravity,  as this could provide insights on  the 4d problem.   The above discussion based on \eqref{CC1} is essentially decoupled from gravity.    
 We thus ignore gravity and focus on defining and calculating $\rhovac$ purely in quantum field theory.      In 2d  the analog of \eqref{CC2} suffers from the same problems,  in that it is also divergent for a free massive particle: 
\beq
\label{CC3}
\rhovac = 2 \int_0^\Lambda  \frac{ dk}{2 \pi} ~ \smallhalf  \sqrt{k^2 + m^2} \approx  ~\frac{\Lambda^2}{4 \pi} + \frac{m^2}{4 \pi} \log ( 2 \Lambda /m) .
\eeq
 In the Concluding Remarks section we return to 4d  and  speculate on experimental data for  $\rhovac$ in the context of real cosmology.    

The main result of this paper is that interactions in QFT can potentially fix  the above problems.   Although the problems with 
equations \eqref{CC2} and \eqref{CC3} do not involve interactions,   the idea is that  proper regularization procedures for interacting QFTs can eliminate divergences in the non-interacting part of  $\rhovac$ in a non-perturbative manner.       In 2d we propose that 
 \beq
 \label{CC4}
 \rhovac = - \frac{m^2}{2 \gcal}
 \eeq
 where $\gcal$ is a dimensionless   interaction coupling which goes to zero in the free field limit,  and $m$ is a fundamental physical mass scale, which we will define in some specific examples in the sequel.    
      That $\gcal$ is a coupling is clear for instance in the sinh-Gordon model (see equation  \eqref{shGvac2}).    Since $\rhovac \propto 1/\gcal$,  this a non-perturbative result.   In the free field limit as 
 $\gcal \to 0$ one recovers $\rhovac \to \infty$ which is consistent with the above discussion of the cosmological constant problem for free theories.     We will present many  illustrative examples.   For the sine-Gordon model,  we see that $\rhovac$ can diverge at special values of the coupling where the theory is not free,  but this arrises due to a kind of 
 ``resonance"  of conformal perturbation theory with $\rhovac$,   as we will explain.        For massive cases,  $m$ is the {\it lightest} mass particle and 
 $\gcal$ will be related to trigonometric sums over certain resonance  pole angles in 2-body  S-matrices.  
 As it turns out $\rhovac$ can be positive,  negative or zero,   depending on the model.    In the sequel we may  refer to   $\rhovac$ in the above 2d equation as the ``cosmological constant",   although this is somewhat of an abuse of terminology since  there is little 4d cosmological about it.   
 
 The word ``mingling" in the title refers to the idea that although $\rhovac$ has low energy implications,   it is inaccessible by perturbative analysis around  the low energy infrared (IR) theory.   It must  be computed in the UV limit,  but even in this limit it is inaccessible by  a perturbative analysis aboout the UV conformal field theory (CFT),  as will be clear in the sequel.   We will actually need an IR regulator corresponding to finite volume to ultimately calculate 
 $\rhovac$.     Thus $\rhovac$ corresponds to an energy scale somewhere between the IR and UV limits which is non-perturbative and difficult to analyze.    The integrability in 2d will allow calculation of $\gcal$ and explain the meaning of $m$.

 We will consider a wide variety of theories that lead to \eqref{CC4}. 
     For simplicity,  we mainly consider examples with only 1 or 2 particles with
 diagonal scattering,  except for the sine-Gordon model which can have many bound states (breathers) and the scattering is generally non-diagonal.    This limitation still allows us to explore several kinds of theories with rather different properties.      In Section II we consider massive theories which are relevant perturbations of a UV conformal field theory,  in particular the sinh and sine Gordon models,  and   integrable  perturbations of the Yang-Lee and 3 state Potts models.   This section relies heavily on  work of Al.  Zamolodchikov \cite{ZamoTBA}  and is essentially a reinterpretation of some of his results,  in particular the $\cbulk$ term and its relation to $\rhovac$. 
 In this context,   although the existence of $\cbulk$ has been recognized and studied,    to our knowledge its interpretation as a 2d analog of a cosmological constant, 
 i.e.  
 consistent with \eqref{CC1},  has not been  discussed in the literature,   although the connection we make is quite straightforward. 
 This is perhaps due to the fact that previous literature was largely focussed on the trace of $T_{\mu\nu}$  rather than its separate energy density and pressure components as in \eqref{Txy}.  
       The next two sections present  more  original results.     
 In Section III we study the irrelevant $T\Tbar$ deformations of an infrared  CFT,  which is a theory of massless particles,  and  may or may not have a UV limit depending on the sign of $\gcal$.      In Section IV we consider massless flows from a UV CFT to a non-trivial IR CFT and present a new general formula for $\rhovac$ in this context;      it appears to differ from  the one known result in the literature of the flow from the tricritical Ising to Ising model by a factor of $2$.    
    In all cases we present several examples.    In the massless cases $m$ in \eqref{CC4} refers to a fundamental energy scale, such as a scale of spontaneous (super) symmetry breaking.      In the Concluding Remarks we speculate on the implications of our results for 4 spacetime dimensions and compare with data on the cosmological constant.

 \def\GammaUV{\Gamma_{\rm UV}}

\section{Massive case}

\subsection{General considerations}

In order to make sense of $\rhovac$ we consider the QFT  in euclidean space,  specifically on a cylinder of infinite length and finite circumference $R$.   
This is an infrared  regularization since it corresponds to a finite volume $R$.    The free energy density takes the form 
$\CF (R) =  E(R) / R $ where $E(R)$ is the ground state energy.    In thermal field theory,   $R=1/T$ where $T$ is the temperature.  
In 2d,  it is conventional to express $E(R)$ in terms of an effective Virasoro central charge $c(mR)$:
\beq
\label{ER}
E(R) = - \frac{\pi}{6} \, \frac{c(mR)}{R} 
\eeq
where at renormalization group fixed points,  which are CFTs,  $c(mR)$ is a constant independent of $R$.   
For  the massive case,   in the above equation $m$ is a physical mass of a particle.   In this euclidean space,  let $t$ denote the spatial coordinate along the circumference and $x$ the coordinate along the length of the cylinder.  
Then standard statistical mechanics implies 
\beq
\label{Txy}
\inv{2 \pi} \langle T_{tt} \rangle =  \frac{d E(R)}{d R} = \CE, ~~~~~~
\inv{2 \pi} \langle T_{xx} \rangle  = \frac{E(R)}{R} = \CF = - p 
\eeq
where $\CE$ is the energy density and $p$ the pressure, and  we have used that $p= - \CF$.  The above implies $\langle T_{\mu\nu} \rangle/2 \pi  =  {\rm diag}  
(\CE, - p)$\footnote{The extra factors of $2 \pi$ are due to the standard CFT normalization for the stress energy tensor in 2d.} \footnote{One can check the signs in 
\eqref{Txy} by considering $\CE$ and $p$ for the CFT of massless particles,   which is the analog of black body radiation in 2d.}. 
Note that $\langle T_{ij} \rangle$ is not proportional to $\delta_{ij}$  in general,   that is to say \eqref{CC1} is violated,  which is obviously due to the $R$ dependence.   
Given the form of $\langle T_{xx} \rangle$,   we define $\rhovac$ via the pressure $p$:
\beq
\label{rhovacTyy}
\rhovac = -\inv{2 \pi} \langle T_{xx} \rangle_0 .
\eeq
In the above equation the $0$ subscript refers to the vacuum expectation value $\langle 0| T_{xx} | 0 \rangle$,   which is the $R$-independent part of  
$\langle T_{xx} \rangle $.    Extracting this $R$-independent contribution is not so straightforward as we will see,  but can be done in the models considered here.

Henceforth let us define the dimensionless quantity $r=mR$. 
At low energies,  i.e. in the IR as $r \to \infty$ one expects $c(r) \to 0$ since the particles are effectively infinitely massive and the IR CFT is empty.    On the other hand,  if the UV theory is defined,  i.e. ``UV complete",  then it can be formulated as a perturbation of a UV CFT with central charge $\cUV$ by a relevant operator of dimension $\GammaUV$.    One then can anticipate that in the UV as $r\to 0$:
\beq
\label{cUVr}
\lim_{r \to 0} c(r) = \cUV + \cbulk + c_{\rm pert}. 
\eeq 
Above,  $c_{\rm pert}$ are terms that can in principle be computed in conformal perturbation theory about the UV,   
and is  an expansion in powers of $r^{2 - \GammaUV}$.    
    What is of interest in the present work is  the $\cbulk \propto r^2$ term which corresponds to
an $R$-independent term in $\langle T_{xx} \rangle$,  which we identify with $\langle 0|T_{xx} | 0 \rangle$.   In other words the $\cbulk$ term implies a vacuum expectation of the 
stress energy tensor that is consistent with a cosmological constant,  i.e. \eqref{CC1}.       It is difficult to compute since it is invisible in the IR where $c(r)$ decreases exponentially, and in the UV it  can mingle  with the $c_{\rm pert}$ terms.      For the moment we    parameterize it  in terms of a constant $b$: ~ 
$\cbulk = - b \,  r^2 /2 \pi$. 
Then based on \eqref{Txy} one has 
\beq
\label{TxyCosmo}
\langle T_{tt} \rangle_0 = \langle T_{xx} \rangle_0 = \frac{\pi}{6} \, b\,  m^2 .
\eeq
It is important to note that this result is of the form $\langle T_{ij} \rangle_0 \propto \delta_{ij}$ which is consistent  with a cosmological constant
\eqref{CC1} where $p = -\CE$;     we mention this fact since  other kinds of calculations  analogous to  \eqref{CC2} can  contradict 
\eqref{CC1}.  See for instance \cite{Martin} and references therein.   
Let us now define $b\equiv 6/\gcal$.     
The equation \eqref{TxyCosmo}  then leads to  the formula in the Introduction
\beq
\label{cbulkrhovac}
\cbulk = - \frac{3 r^2}{ \pi \gcal} ~~~~\Longrightarrow ~~~\rhovac = - \frac{m^2}{2\gcal}. 
\eeq
For CFTs,   $\rhovac =0$.      We emphasize that we view the introduction of $r$ and $c(r)$ as essentially a tool to extract 
$\rhovac$,   however in principle it should be possible to define $\rhovac$ directly from the zero temperature S-matrix,   but  we do not attempt this here.

There is an important caveat to the potential validity of \eqref{cbulkrhovac}.    As stated above,   $\cpert$ is expected to be
an expansion in $r^{2-\GammaUV}$.    Due to conformal bootstrap selection rules,  i.e. the OPE structure,  $\cpert$ is typically a sum of terms
proportional to $r^{2n(2 - \GammaUV)}$ with $n=1,2,3, \ldots$.     
Now if $\GammaUV <1$,  then $\cbulk$ is the leading term as $r \to 0$,   and is more easily extracted from $c(r)$.   
On the other hand,   for $1 <\GammaUV < 2$,   the terms in $\cbulk$ and $\cpert$ are more entangled since $\cbulk$ is no longer the leading term and $\cbulk$ is difficult to extract numerically.      More importantly,   there exists the possibility of a resonance in conformal perturbation theory where one term in $\cpert$ goes exactly as $r^2$.      This occurs for rational values of $\GammaUV$: 
\beq
\label{resonance}
\GammaUV = 2 - 1/n,  ~~~~n = 1,2,3,\ldots ~~~~({\rm resonance }).
\eeq
This resonance phenomenon was understood by Al. Zamolodchikov for ``RSOS" perturbations of minimal models of CFT \cite{ZamoRSOS}.  
If this occurs then $\rhovac$ may still diverge due to divergent terms in $\cpert$.    We will present a precise example of this phenomenon for the sine-Gordon theory below.    
However we wish to emphasize that this resonance phenomenon is specific to the TBA at finite $R$ where the term of interest $\cbulk$ can mix with $\cpert$,  whereas $\rhovac$ has a  meaning in the pure QFT without this complicated technicality;  i.e. the TBA is used here as a tool to extract the independently meaningful $\rhovac$.

Henceforth we present results for $\gcal$  (rather than $b$) for various models based on the definition \eqref{cbulkrhovac}.
Below we will show that $\gcal$ is essentially an effective coupling of the underlying QFT.     For many specific integrable models 
this coupling is a fixed  number with no variable parameters,  which we will calculate in some examples,   although they are often related to models with variable couplings but at fixed  rational values.          For other models,   such as the sinh-Gordon and sine-Gordon models 
considered below,   it does indeed depend on 
a continuously variable coupling.  See  for instance equations \eqref{shGrhovac} and \eqref{rhovacSineG}.

\subsection{Thermodynamic Bethe Ansatz for Integrable theories}

For simplicity let us first assume the spectrum consists of a single massive particle of mass $m$ and the QFT is integrable.   
The  energy and momentum of this particle can be parametrized  in terms of a rapidity $\theta$: 
\beq
\label{rapid}
E = m \cosh \theta,~~~~~ p = m \sinh \theta.
\eeq
Al. Zamolodchikov obtained seminal results for $E(R)$ based on the Thermodynamic Bethe Ansatz (TBA) \cite{ZamoTBA} by extending results of Yang and Yang \cite{YangYang} to relativistic theories.  The rest of this section relies primarily  on Zamolodchikov's analysis,  
thus let us present these TBA equations.        
Let $S(\theta)$ denote the 2-body  S-matrix and define the kernel 
\beq
\label{kernel}
G(\theta) =  -i \d_\theta \log S(\theta) .
\eeq  
Unitarity $S(\theta) S(-\theta) =1$ implies $G(-\theta) = G(\theta)$.
Introduce a so-called pseudo-energy $\vep (\theta )$ which is a solution to the integral equation:
\beq
\label{TBAmass}
\vep (\theta) = r  \, \cosh \theta  -  G \star \log\( 1+ e^{-\vep} \)
\eeq
where we  have defined 
$\( G \star f \) (\theta) = \int_{-\infty}^\infty \frac{d\theta'}{2\pi} \, G(\theta - \theta') f(\theta') $
for an arbitrary function $f(\theta)$.  
Then the $r$ dependent central charge in \eqref{ER} is given by 
\beq
\label{cmass}
c(r) =  \frac{3}{\pi^2} \int_{-\infty}^\infty d\theta ~ (r \cosh \theta) \, \log \( 1+ e^{- \vep (\theta)} \) .
\eeq

As expected,  in the IR limit $r\to \infty$,   $c(r) \to 0$ exponentially as $e^{-mr}$.   This is easily seen from the TBA equation as the first iterative 
correction to $\vep (\theta)$ comes from replacing $\vep (\theta)$ with $r \cosh \theta$ in the $G\star \log $ term on the RHS of \eqref{TBAmass},  
which leads to various Bessel functions.     The UV limit $r \to 0$ is much more delicate.     
In the range 
$ -\log (2/r) \ll \theta \ll \log (2/r) $,  $\vep (\theta) \approx \vep_0$ where $\vep_0$ is a constant satisfying the ``plateaux" equation:
\beq
\label{plateaux}
\vep_0 = -\, k  \,\log \( 1+ e^{- \vep_0} \),  ~~~~~ {\rm with}  ~~~~~k  = \int_{-\infty}^\infty \frac{d \theta}{2 \pi} \, G(\theta) .
\eeq
The UV central charge $\cUV$ is a function of $\vep_0$ which can be expressed in terms of the Roger's dilogarithm.     For our purposes the TBA also gives 
$\cbulk$ as discovered in \cite{ZamoTBA},   thus let us present this result.   If the kernel behaves as 
\beq
\label{gfromG}
G(\theta) = - \gcal \, e^{- |\theta|} + O(e^{-2 |\theta|})
\eeq
then 
\beq
\label{cbulkb}
\cbulk = - \frac{ 3  r^2}{\pi \gcal}     
\eeq
which implies the result \eqref{cbulkrhovac}.  

The constant $\gcal$ is related to certain  resonance angles  in the S-matrix $S(\theta)$.   The basic building blocks of $S(\theta)$ 
are factors of $f_\alpha (\theta)$: 
\beq
\label{Sf}
S (\theta) = \prod_{\alpha \in \CA}  f_\alpha (\theta)
\eeq
where
\beq
\label{falpha}
f_\alpha (\theta) = \frac{ \sinh \smallhalf \( \theta + i \pi \alpha \)}{\sinh \smallhalf \( \theta - i \pi \alpha \)} 
\eeq
and $\CA$ is a finite set of $\alpha$'s.   
One has 
\beq 
\label{fkernel} 
-i \d_\theta \log  f_\alpha (\theta) = - \frac{\sin \pi \alpha}{\cosh \theta - \cos \pi \alpha }
\eeq
which leads to 
\beq
\label{galpha}
\gcal =2  \sum_{\alpha \in \CA}  \sin \pi \alpha .
\eeq
The angles $\pi \alpha$ correspond to poles in the complex $\theta$ plane for the S-matrices.

A remarkable result is that for many particles of mass $m_a$,  $a=1,2,3, \ldots$  with diagonal scattering,   due to the S-matrix bootstrap,  the formula \eqref{cbulkb} holds with $m=m_1$ the {\it lightest} particle and 
$\gcal$ given by the sum over resonance angles of the S-matrix  $S_{11} (\theta) $ for  the scattering of $m_1$ with itself.  
Specifically,   if $G_{11} = -i \d_\theta \log S_{11}  (\theta)$,  then $\gcal$  is defined by  $G_{11}  (\theta) = - \gcal\,   e^{-|\theta|} + \ldots$  as in
\eqref{gfromG}.    We thus arrive at 
\beq
\label{rhovac11}
\rhovac = - \frac{m_1^2} {2 \, \gcal} ~.
\eeq
The derivation of these results were well explained in \cite{KlassenMelzer,MussardoBook} thus we don't repeat  the derivation here.
Generalizations of the above formula for massless theories will be derived below,  and this will give the reader an idea of how  
formulas such as \eqref{rhovac11} are obtained.   
We point out that the above formula \eqref{rhovac11} has not been proven in complete generality;   it has  been proven for diagonal S-matrices under the assumption that there are no $r^2$ terms in $c_{\rm pert}$,  namely that the resonance     condition \eqref{resonance}  does not occur.    Below we will consider the sine-Gordon example where it does occur in a regime of the coupling.

\subsection{Examples}

We present 4  examples to illustrate that the cosmological constant $\rhovac$ can be positive,  negative, or zero,   and this distinction is not obviously related to the unitarity of the model,  nor does it require spontaneous symmetry breaking.  
    The first example is the sinh-Gordon model,  and the second is the sine-Gordon theory. 
The next two examples were already considered in \cite{ZamoTBA},  but are included for comparison and illustration.   
In two of  these examples the S-matrix is a product of two $f_\alpha$ factors:
\beq
\label{Falpha}
 F_\alpha (\theta) = f_\alpha (\theta) f_\alpha ( i \pi - \theta) = \frac{ \sinh \theta + i \sin \pi \alpha }{\sinh \theta - i \sin \pi \alpha} , 
\eeq
where the  kernel is 
\beq
\label{Fkernel}
 - i \d_\theta \log F_\alpha (\theta) = - \frac{2 \cosh \theta \sin \pi \alpha}{\cosh^2 \theta - \cos^2 \pi \alpha }. 
 \eeq
 Thus for each $F_\alpha$ factor 
\beq
\label{gcalF}
\gcal = 4 \sin \pi \alpha .
\eeq

\bigskip

\noindent $\bullet$  {\bf Sinh-Gordon model.} ~ This is a unitary theory with $\cUV = 1$.   
It is defined by the following action for a single scalar field \footnote{Here $b$ is not to be confused with $b$ in \eqref{TxyCosmo}.} :
\beq
\label{shGaction}
\CS = \int d^2 x \( \inv{8 \pi} (\d_\mu \phi \, \d^\mu \phi ) + 2 \mu \cosh ( \sqrt{2} \, b \phi ) \) .
\eeq
Expanding out the $\cosh$ it is clear that $b$ is a coupling.   
There are no bound states and the spectrum consists of a single particle of mass $m$ corresponding to the field $\phi$ itself,  which qualifies it as perhaps the simplest integrable QFT. 
The $\cosh$ perturbation has scaling dimension $\GammaUV = - 2 b^2$,   which is strongly relevant,  and we don't expect the 
phenomenon of resonance \eqref{resonance} to occur since $\GammaUV <1$. 
The S-matrix is very well known,   as are it's form factors (for a comprehensive treatment we refer to \cite{MussardoBook} and references 
therein).  The S-matrix is 
\beq
\label{shGSmatrix}
S (\theta) = F_{-\gamma} (\theta) , ~~~~~ \gamma = b^2/(1 + b^2) .
\eeq
It is invariant under the weak strong duality $b \to 1/b$.   
This gives $\gcal = - 4 \sin \pi \gamma$, and the cosmological constant is {\it positive}:
\beq
\label{shGrhovac}
\rhovac =  \frac{m^2}{8 \sin \pi \gamma}  .
\eeq
For small coupling $b$ 
\beq
\label{shGvac2}
\rhovac \approx \frac{m^2}{8 \pi b^2}  .
\eeq
This positivity is typical of spontaneous symmetry breaking,   however this is not known to occur in this model.   
As anticipated in the Introduction,   $\rhovac$ diverges in the free field limit $b\to 0$,  and also as $b \to \infty$ due to the duality $b \to 1/b$.     
This result makes it clear that $\rhovac$ is highly  non-perturbative.   It would be very interesting to derive this result 
in a semi-classical analysis for small $b$,  however this is beyond the scope of this paper \footnote{Note added:  Some results on this will be presented in \cite{AL4d}.}.

\def\betahat{\hat{\beta}}

\bigskip
\noindent $\bullet$  {\bf Sine-Gordon model.} ~ This is a unitary theory with $\cUV = 1$.   
It can be  defined by the following action for a single scalar field:
\beq
\label{shGaction}
\CS = \int d^2 x \( \inv{8 \pi} (\d_\mu \phi \, \d^\mu \phi ) + 2 \mu \cos  ( \sqrt{2} \, \betahat  \phi ) \) .
\eeq
The $\cos$ term has dimension $\GammaUV = 2 \betahat^2$,   thus it is relevant for $0 < \betahat^2 < 1$. 
 
Compared to the sinh-Gordon model,   this model has a much more complex spectrum of particles. 
For any $\betahat$ there is a soliton and anti-soliton of mass $m_s$,   and these are the only particles for $\betahat^2 > 1/2$.
In the massive Thirring description these are charged fermions.   At the coupling $\betahat = 1/\sqrt{2}$ the theory is a free charged Dirac fermion.   
The marginal point occurs at $\betahat =1$.     For $\betahat< 1/\sqrt{2}$ the interactions are attractive and the solitons form bound states called ``breathers".      For generic $\betahat$,    the scattering is non-diagonal and the formulas in Section II do not  necessarily apply.    However
at the couplings 
\beq
\label{betahatfreflect}
\betahat^2 = 1/j,   ~~~~{\rm for} ~~  j=2,3, \ldots
\eeq
 then there are $j-2$ breathers and the scattering is diagonal.   These points are referred to as ``reflectionless" due to the diagonal scattering. 
    The soliton is obviously the lightest mass particle.      Letting $s, \bar{s}$ denote the soliton,  anti-soliton,    it's S-matrix is well-known \cite{ZamoZamo0}:
\beq
\label{reflectionlessS}
S_{s \bar{s}} = \prod_{k=1}^{j-2} f_{k/(j-1)} (\theta) .
\eeq
The cosmological constant can then be computed from \eqref{rhovac11}.    
This gives $\gcal = 2 \sum_{k=1}^{j-2} \sin ( \pi k/(j-1) ) = 2 \cot ( \pi/(2j -2) )$.    Expressing $j$ in terms of $\betahat$,   
 then at the reflectionless points the cosmological constant is  $ {\it negative}$   for $\betahat^2 < 1/2$:
\beq
\label{rhovacSineG}
\rhovac =  - \frac{m_s^2}{4} \, \tan \( \frac{ \pi \betahat^2}{2 ( 1- \betahat^2 )} \) .
\eeq
The above formula was first obtained by different,  somewhat more difficult methods which do not rely on analytically continuing the reflectionless points   in two different ways \cite{DestriDeVega,ZamoSineG}: from the light-cone lattice regularization and also by analytic continuation from  the sinh-Gordon theory $b \to i \betahat$.     For the latter one must first express the result in terms of
the mass of the first breather rather than the soliton mass $m_s$.\footnote{The result in \cite{DestriDeVega} seems to have missed a factor of $4$ compared to the result in  \cite{ZamoSineG},  where the latter agrees with \eqref{rhovacSineG}. }    The generalization of this sinh-Gordon/sine-Gordon correspondence between breathers and solitons to affine Toda theories was  understood in \cite{MacKay,Takacs}.

The above formula has several  very interesting  and illustrative features.     Consider first the true reflectionless points in the region 
$0 < \betahat^2 < 1/2$  where the above formula was derived.   
 First it is expected to be strictly valid for $\betahat^2 < 1/2$ where 
$\GammaUV <1$. 
 It diverges precisely at the free-fermion point $\betahat^2 = 1/2$;   this is consistent with the remarks in the Introduction that $\rhovac$ diverges for free theories,  but interactions can render it finite.    
However it does not diverge as $\betahat \to 0$,   since it is not quite a free theory due to the infinite number of breathers 
and $\betahat \to 0$ is a delicate limit. 
In fact,  contrary to the sinh-Gordon theory,   $\rhovac =0$ when $\betahat =0$.    
This also does not contradict the statements in the Introduction.

Let us assume that we can analytically continue  the formula \eqref{rhovacSineG} to $\betahat^2 > 1/2$.    It is not  a  priori clear this is allowed, since in this region  there are no breathers and the scattering is non-diagonal,  thus  the analysis of Section II  does not necessarily apply.    We now argue that this formula is indeed valid in this region with the proper interpretation,   and has some physically meaningful properties that can be explained by previous  older  work. 
First of all,  for irrational $\betahat$,   $\rhovac$  in \eqref{rhovacSineG} is finite everywhere.    Thus one needs to explain its zeros and infinities.   Let us first consider the zeros.   
In \cite{BernardLeClair} it was shown that the sine-Gordon theory has non-local conserved charges $Q_\pm, \bar{Q}_\pm$  that satisfy the quantum affine algebra
$\CU_q (\hat{{\rm su} (2)}) $ where $q= e^{- i \pi/\betahat^2}$.    When
\beq
\label{betap}
 \betahat^2 = p/(p+1),  
 \eeq
 where $p \geq 1$ is an integer,  $q= - e^{-i \pi /p}$ is a root of unity.    In \cite{Vafa} it was shown that when $p$ is {\it even},   the quantum affine algebra corresponds to a fractional supersymmetry,  
where the conserved charges have spin $\pm 1/p$.     For $\betahat^2 = 2/3$,  i.e.  $p=2$,  the quantum affine symmetry is equivalent to 
$\CN =2$ supersymmetry and is a natural generalization of it  for higher $p$.       For general $p$,   ``fractional supersymmetry" refers to the fact that mathematically the hamiltonian $H$  is in the center of the quantum affine algebra,  namely $[Q, H]=0$ for all the non-local charges 
$Q$.    For instance,   for $p=4$,    $Q^4 = H$,   where $Q^4 = Q_+ Q_- Q_+ Q_- + ....$ is quartic  in the conserved charges $Q_\pm$ 
(see \cite{Vafa} for the explicit expression,   which we do not need here).      The 
main point  we wish to make is that this fractional supersymmetry,  which is unbroken, $Q|0\rangle =0$,   implies that $\rhovac =0$ at these points,  for the same reason that ordinary $\CN=2$ supersymmetry implies $\rhovac =0$ in general,  and  \eqref{rhovacSineG} indeed shows that it 
does vanish, i.e. $\rhovac =0$,  for $p=2$  and higher even $p$.\footnote{H.  Saleur once pointed out to us that the vanishing of the free energy density  at these even $p$ points is consistent with lattice results obtained by 
Baxter for the XYZ model \cite{Baxter},  however we could not find the exact reference.}  
This supports the analytic continuation of the formula \eqref{rhovacSineG},  at least to  these $p$ even  integer points\footnote{In \cite{ZamoRSOS} it was noted  that the vacuum energy vanishes
for  integrable perturbations of  some of the lowest minimal models of CFT with $c<1$ due to a fractional supersymmetry,  however this is not really the same as the vanishing
of $\rhovac$ in the (unrestricted) sine-Gordon model itself  with $c=1$ based on the fractional supersymmetries in \cite{Vafa},  but rather due to the related residual symmetries upon restriction of the sine-Gordon model to perturbations of the minimal models \cite{Residual}. } .

Let us now turn to the divergences  in \eqref{rhovacSineG}.   Namely   
   $\rhovac$ diverges at the {\it odd}  $p$ points when $\betahat^2$ is given by \eqref{betap}.    Assuming  this analytic continuation is valid,  it appears possible that 
  $\rhovac$ can  also diverge at  special non-free points $p = {\rm odd}$.     However in this case this divergence  can be explained by the resonance phenomenon captured in the condition \eqref{resonance},  and thus come from $\cpert$.        Here this resonance  condition is simply $\betahat^2 = p/(p+1)$ where $p= 2n -1$ is odd.  

In summary,   $\rhovac$ is negative and finite  in the region below the free fermion point $\betahat^2 = 1/2$,  where it diverges. 
   Assuming analytic continuation in $\betahat$  is possible, which we argued that it is,   then 
above this point  $\rhovac$ has an infinite number of special points where it is either zero or $\pm \infty$,  otherwise it is finite.   
 Thus as $\betahat$ approaches the marginal point $\betahat = 1$,  $\rhovac$ has an infinite number of oscillations between $\pm \infty$,   
 which we attribute to the perturbative $\cpert$,  and not the cosmological constant term $\cbulk$.

\bigskip

\noindent $\bullet$   {\bf Yang-Lee  model.} ~ Here the UV CFT is non-unitary with Virasoro central charge $c=-22/5$.   
The TBA computes the effective central charge in the UV $
c_{\rm eff} = c - 12 d_0$, 
where $d_0$ is the lowest scaling dimension of primary fields,   which here is $d_0 = -2/5$.    Thus $c_{\rm eff} = 2/5$.  
The  S-matrix is 
\beq
\label{SLeeYang}
S (\theta) = F_{1/3} (\theta)  .
\eeq
Thus $\gcal = 4 \sin (\pi/3) $,  and one has 
\beq
\label{LeeYangrhovac}
\rhovac = -\frac{m^2}{4 \sqrt{3}}.
\eeq
This corresponds to a {\it negative}  cosmological constant.  

\bigskip

\noindent $\bullet$  {\bf Non critical 3-state Potts model.}  ~ This is a perturbation of the unitary critical 3-state Potts model with central charge $\cUV=4/5$ by an operator of dimension $\GammaUV = 2/5$,   which is a temperature perturbation.        Here there are two particles of equal mass.   Labeling them $1,2$ the
S-matrices are 
\beq
\label{Potts}
S_{11} (\theta) = f_{1/3} (\theta), ~~~~~ S_{12} (\theta) = f_{1/6} (\theta) .
\eeq
This implies $\gcal = \sqrt{3}$,  and once again this gives a  negative  cosmological constant:
\beq
\label{Pottsrhovac} 
\rhovac = - \frac{m^2}{2 \sqrt{3}} .
\eeq
Thus it appears that the sign of the cosmological constant does not depend on unitarity.   
\bigskip

\section{First  Massless case:  Pure $T\Tbar$}

In this section we consider so-called $T\Tbar$  deformations of CFT's.    
There is a large literature on this subject,   however we mainly rely on 
\cite{Tateo1, SmirnovZ},   and refer to these papers for additional references.

Every theory has a conserved stress-energy tensor, 
\beq
\label{Conv1}
T =-2 \pi \, T_{zz} , ~~~\Tbar = -2 \pi \, T_{\zbar\zbar}, ~~~ \Theta = 2 \pi\,  T_{z\zbar} , 
\eeq
where $z=x+i y$, $\zbar = x-i y$ are euclidean light-cone coordinates and $\Theta$ is its trace.     
With these components one can define the operator
\beq
\label{Conv2}
T \Tbar = 4 \pi^2 \( T_{zz} T_{\zbar\zbar} - (T_{z \zbar})^2 \) . 
\eeq
For a CFT,   $\Theta =0$.    Let $\CS_{\rm cft}$ formally denote the action for a CFT
and consider a perturbation by $T\Tbar$:
\beq
\label{Conv3}
 \CS =  \CS_{\rm cft }  +  \frac{\alpha}{\pi^2} \int d^2 x \, T\Tbar .
\eeq
Since $T\Tbar$ is an irrelevant operator,   the CFT is a description of the IR limit,  and this must be a massless theory.      Let us denote the central charge of this CFT as  $\cIR$.  
Being an irrelevant perturbation,  the above action is not renormalizable since additional higher irrelevant operators are generated in perturbation theory.    Thus to be more precise    
\beq
\label{CS2}
\frac{d \CS_\alpha }{d \alpha} = \inv{\pi^2} \int d^2 x \, \, (T\Tbar)_\alpha 
\eeq 
where $(T\Tbar)_\alpha$ signifies the  expectation value of the operator $T\Tbar$ at stages of the renormalization group flow generated by $\alpha$ \cite{Tateo1,SmirnovZ}.  
 The function $\CS_\alpha$ can be interpreted as an effective action.    
 Since $\alpha$ has units of inverse mass squared,  we define the dimensionless coupling 
 \beq
 \label{galpha} 
 g= -\alpha \, m^2 
 \eeq
 where $m$ is a fundamental  mass scale indicative of the breaking of conformal invariance.         
 
The main interesting aspect of these $T\Tbar$ deformations is that the ground state energy $E(R,\alpha)$ is easily related to $E(R)$ with $\alpha =0$.    
 One way to formulate  this is based on the inviscid Burgers differential equation satisfied by $E(R)$;  ultimately it depends on the operator algebra 
 of $T_{\mu\nu}$.    In any case the result is well explained in the original works \cite{Tateo1,SmirnovZ},  and is rather simple:
\beq
\label{cTT}
c(r) = \frac{2 \cIR} {1 + \sqrt{ 1-   \frac{2 \pi g}{3r^2}  \cIR   }} . 
\eeq

Since this is a massless theory,   in the IR limit  $r\to \infty$,  $c(r)$ is not zero,  namely it is $\cIR$ in this limit.    One finds 
\beq
\label{cTTrInfinity}
\lim_{r \to \infty} c(r) = \cIR + \frac{ \cIR^2 \pi g}{6 r^2 } + \frac{\cIR^3 \pi^2 g^2}{18 r^4} + O(g^3/r^6 ) .
\eeq
In the  UV limit  $r \to 0$ on the other hand one finds 
\beq
\label{cTTr0} 
\lim_{r \to 0} c(r) = - \sqrt{- \frac{6 \cIR} {g\pi} }  ~ r + \frac{3 r^2}{ \pi g} - \inv{ \sqrt{-\cIR} } \( \frac{3}{2 \pi g} \)^{3/2} \, r^3 + O(r^5) .
\eeq

Depending on the sign of $\cIR/g$ the UV limit may or may  not be defined since the square root could be imaginary.    The leading  linear in $r$ term should be interpreted as a contribution to $c_{\rm pert}$  since   perturbation around $\cUV$ indicates that $\GammaUV =1$ \cite{ALcompletion}.  For both $\cIR$ and $g$ positive the UV limit is not defined,   which has been interpreted as a Hagedorn transition in a string theory context \cite{Dubovsky,Verlinde}.     For further analysis of these transitions in some simple cases see \cite{Camilo}.     On the other hand if 
$\cIR/g$ is negative  the UV limit is well defined and $\cUV =0$.   

What is important here is that  whatever the UV limit is,  one can easily isolate the term of interest corresponding to $\rhovac$,  which is actually independent of $\cIR$:   
$\cbulk = 3 r^2/ \pi g$.   In the next section we will consider massless theories with a UV completion.   Comparison with \eqref{cbulkb},   one identifies $\gcal = -g$  which gives
\beq
\label{TTrhovac} 
\rhovac =  \frac{m^2}{2g} .
\eeq
This cosmological constant can be positive or negative depending on the sign of $g$.      For the most interesting case of both $\cIR$ and $g$ positive,  it is positive.

\def\LL{{\rm LL}}
\def\RR{{\rm RR}}
\def\LR{{\rm LR}}
\def\RL{{\rm RL}}

\section{Massless case for UV complete theories.} 

\subsection{Generalities}

In the present context,   massless theories are either CFT's themselves,  or  renormalization group 
flows between relevant perturbations of a UV CFT and 
a non-trivial IR CFT.      If the theory is a CFT,   then $\cbulk$ and $\rhovac=0$.   For massless flows 
 we derive formulas analogous to \eqref{cbulkrhovac} which were not previously known in general.   For pure $T\Tbar$ as we reviewed in the last section,  if for instance $\cIR >0$
and $g>0$ then the theory cannot be extended to arbitrarily short length scales without first encountering a Hagedorn transition.     However,  
if one adds additional higher spin versions of $T\Tbar$,  denoted $[T\Tbar]_s$ for $s$ a positive integer, 
then integrability is maintained \cite{SmirnovZ},  and if one fine tunes the higher couplings one can propose UV completions of the original
$T\Tbar$ deformation of the CFT with $\cIR$ \cite{ALcompletion}.      The present goal is to calculate $\rhovac$ in the form \eqref{cbulkrhovac} for such a class of theories,  which amounts to calculating $\gcal$.    
 
We follow the  scattering description of these massless theories formulated  more generally in \cite{ZZ}.  
In this situation one must distinguish Left/Right movers (L/R),   which can also be parameterized by a rapidity $\theta$:
\barray
\label{movers} 
{\rm right ~ movers:} ~~~~~ E &=&  p= \tfrac{m}{2} e^{\theta} \cr
{\rm left  ~ movers:} ~~~~~ E &=& - p= \tfrac{m}{2} e^{-\theta}  .
\earray
Then the S-matrix elements satisfy 
\beq
\label{Sss}
S_\LL (\theta) = S_\RR (\theta), ~~~~~~~ S_\RL (\theta) = S_\LR (-\theta).    
\eeq
If $S_\RL = 1$,   then the theory is conformally invariant and $S_\LL, S_\RR$ provide a scattering description of the IR CFT. 
On the other hand $S_\RL \neq 1$ breaks the conformal invariance,   and the mass $m$ in \eqref{movers} is a characteristic scale of 
conformal symmetry breaking.

For simplicity let us  first assume $S_\LL = S_\RR = -1$,  appropriate to a fermionic description,   and the spectrum consists of a single particle.     
The TBA involves 
the kernels which follow from these S-matrices, $G_\RL (\theta) = G_\LR(-\theta) =  -i \d_\theta \log S_{RL}(\theta)$.  
This leads to the now coupled TBA equations for pseudo-energies $\vep_R (\theta) = \vep_L (-\theta)$:   
\barray
\vep_R (\theta) &=&  \tfrac{r}{2} e^{\theta} - \,G_{RL} \star \log \( 1+ e^{-\vep_L } \)    \cr
\vep_L (\theta) &=&  \tfrac{r}{2} e^{- \theta} -  \,  G_{LR} \star \log \( 1+ e^{-\vep_R } \)  .
\label{TBAmassless2}
\earray
Finally,
\beq
c(r) = c_L (r) + c_R (r) = 2 c_R (r) = 2 c_L (r) 
\eeq
where 
\beq
\label{cLR}
c_R (r)  =    \frac{3}{ \pi^2} \,  \int_{-\infty}^\infty d\theta  \, \( \tfrac{r}{2} \, e^{\theta} \) \, \log\( 1 + \, e^{-\vep_R (\theta) } \)   .
\eeq

Let us first explain how to obtain the pure $T\Tbar$ result in \eqref{TTrhovac} from the  above TBA.  
The kernels can be obtained from the massless limit of a CDD factor $S_{\rm cdd} (\theta ) = e^{i g \sinh \theta}$,   which formally requires
its factorization \cite{ALcompletion}.     The result is 
\beq
\label{GLR} 
G_\RL  (\theta) = G_\LR  (-\theta) =  g  \, e^\theta /2
\eeq
from which one can derive \eqref{cTT}.  
Let us thus define $\gcal$ as follows:
\beq
\label{gcalfromGLR}
G_\LR (\theta) =  - \gcal \, e^{-\theta} /2 , ~~~~~~ \gcal = -g .
\eeq
We can express the result \eqref{TTrhovac} in terms of kernel parameters $\gcal$ as follows: 
 \beq
 \label{rhovacmassless}
 \rhovac = \frac{m^2}{2 \gcal}  
 \eeq
 as in \eqref{cbulkrhovac}.

We now briefly show how  the above formula \eqref{rhovacmassless} is more generally valid,  where $\gcal$ is defined in a very similar manner to the massive case,  namely $\gcal$ is defined by 
\beq
\label{GLRgcal}
\lim_{\theta \to \infty} G_\LR (\theta) = - \gcal \, e^{-\theta} /2 + O(e^{-2 \theta}). 
\eeq
Again,  to simplify things we assume a single particle with only $S_\RL, S_\LR$ non-trivial.   In order to dispense of the $\cUV$ term
we start with 
\barray
\label{cprime}
\frac{c'}{r} \equiv \inv{r} \, \d_r c(r) &=& - \frac{3}{\pi^2} \int_{-\infty}^\infty d \theta \, e^\theta \( \inv{ e^{\vep_R } +1} \) \( \d_r - \inv{r} \d_\theta \) \vep_R 
\\
 &=& - \frac{3}{\pi^2} \int_{-\infty}^\infty d \theta \, e^{-\theta}  \( \inv{ e^{\vep_L } +1} \) \( \d_r + \inv{r} \d_\theta \) \vep_L
 \earray 
which can be derived just using a few integrations by parts.    Now $c' /r$ still involves a mingling of $\cbulk$ and $c_{\rm pert}$,  but it is still 
possible to extract the $\cbulk$ term as follows.       
In the equation \eqref{cprime}    the $\inv{r} \d_\theta$ term dominates as $r\to 0$,  and again using an integration by parts one can show
\beq
\label{MasslessTBA1}
\( G_{RL} \star \log (1 + e^{-\vep_L } ) \) (\theta) \approx -  \gcal \( \frac{r e^\theta}{2} \) \( \frac{ \pi}{6} \) \, \frac{c'}{r} .
\eeq
On the other hand,  using \eqref{TBAmassless2} for $\vep_R$,  and matching powers of $e^\theta$ 
implies 
$\frac{c'}{r} \sim   -  \frac{6}{\pi \gcal}$ .
This implies the equation \eqref{cbulkrhovac}  with 
\beq
\label{cbulkLess}
\cbulk  = - \frac{3 r^2}{  \pi \gcal} . 
\eeq

\subsection{Two examples with spontaneous supersymmetry breaking and a massless Majorana fermion}

The critical Ising model has a description in terms of a free Majorana fermion with  Virasoro central charge $c = 1/2$.   
Assuming this spectrum,  integrable  $T\Tbar$ deformations with a UV completion were  classified in \cite{ALcompletion} and only 
two possibilities were found with $\cUV = \tfrac{7}{10}$ or $\tfrac{3}{2}$.    Both of these UV theories have a $\CN  =1 $ supersymmetry 
and the massless Majorana particle can be interpreted as a goldstino arising from spontaneous supersymmetry breaking 
\cite{ZamoFlow,AhnZamo}.       There also exists a scattering description of the Ising model based on the Lie algebra $E_8$ and there are additional possible UV completions with 8 massless particles \cite{Changrim}.   For instance S-matrices for these particles 
were proposed that describe a flow from $\cUV = \tfrac{21}{22}$ to $\cIR = \smallhalf$,  which also arrives to Ising via $T\Tbar$.   These S-matrices have many more $f_\alpha$ factors.
Here we focus on the simpler case of a single massless Majorana fermion.     

\bigskip

\noindent $\bullet$  {\bf Tricritical Ising to Ising.}     This is a  renormalization group flow from $\cUV = \tfrac{7}{10}$  to $\cIR = \smallhalf$ \cite{ZamoFlow}.    The $S_{\LR}$, 
$S_{\RL}$  scattering were proposed in \cite{ZamoFlow},   $S_{\LR} = - \tanh \smallhalf (\theta - i \pi/2) $,   and lead to the following kernels:    
\beq
\label{TIM1}
G_{LR} (\theta) = G_{RL} (-\theta) = \inv{\cosh \theta} .
\eeq
Thus $\gcal = -4$ and the cosmological constant is positive:\footnote{The work \cite{ZamoFlow} 
implicitly writes  a $\cbulk$ term that implies $2$ times the result \eqref{TIM2},  namely $\gcal = -2$ rather than $-4$.
   Since he provides little explanation, 
we don't know the origin of this discrepancy.  However we can point out that $\gcal = -2$ is appropriate to the {\it massive}  case with the
kernel \eqref{TIM1}.}
\beq
\label{TIM2} 
\rhovac = \frac{m^2}{8} .
\eeq

\bigskip

\noindent $\bullet$  {\bf Susy sinh-Gordon to  Ising.}      The UV theory consists of a free boson and a Majorana fermion with $\cUV = \tfrac{3}{2}$.  
Again the flow to the IR leaves behind a massless Majorana fermion goldstino \cite{AhnZamo}.     Here the kernel is twice \eqref{TIM1}  
 at the self-dual point of the susy sinh-Gordon model 
\cite{ALcompletion}.    
Thus $\gcal = -8$ and 
\beq
\label{susyShrhovac}
\rhovac =  \frac{m^2}{16} .
\eeq

\section{Concluding remarks}

We have shown that the analog of the cosmological constant $\rhovac$ can be properly defined in two spacetime dimensions 
for a variety of  types of integrable quantum field theories.     This requires interactions and is non-perturbative in that $\rhovac \propto 1/\gcal$ where 
$\gcal$ is a generalized coupling which is generally zero in the free field limit.   It can be positive, negative or zero depending on the model.    The main tools we used to arrive at this conclusion was integrability with a finite volume $R$,    however it is likely this idea 
is more generally valid.     In fact the value of $\rhovac$ ultimately depends only on properties of the  zero temperature S-matrix,    thus it should be possible to understand $\rhovac$ without the  regulator $R$, in particular without the TBA in 2d.       The work \cite{LecMuss} on one-point functions at finite temperature $T$ suggests that this may be  possible by using form-factors.

Since the original motivation for this work was the observed small positive cosmological constant in 4 spacetime dimensions,   let us speculate on the so-called ``cosmological constant problem" \cite{Weinberg} based on the above results.    Although 2d and 4d physics are very different in details and complexity,   general concepts such as the renormalization group,  are valid in all dimensions,   and one may consider  that conceptual foundations for the computation of $\rhovac$  is in the same category.    Comparing equations \eqref{CC2},\eqref{CC3}  and taking into account 
\eqref{rhovac11},  along with differences of sign in 2d verses 4d,  leads us to propose that in 4d,  restoring factors of $\hbar$ and $c$,  
\beq
\label{rhovac4d}
\rhovac =     \frac{m_1^4}{2 \gcal}  \, \frac{c^5}{ \hbar^3 }
\eeq
where $m_1$ is the lightest mass particle and $\gcal$ is a  positive or negative  dimensionless  interaction coupling based on the S-matrix  for the scattering of this particle with itself.     Of course,   at this point we don't have a detailed  proposal for calculating  $\gcal$,  unlike the 2d models studied above;  the 4d case is much harder without the integrability\footnote{Note added:  some $4d$ results are forthcoming \cite{AL4d}.}.     However based on analogy with the 2d case,   $\gcal$ should be related to resonance poles of the S-matrix for the scattering of mass  $m_1$ with itself.    
    Nevertheless we can at least  check if this idea is at all reasonable.   $\rhovac$ is an energy per volume.    
Based on astronomical observations \cite{WMAP} 
\beq
\label{rhovacexp}
\rhovac \approx 10^{-9} ~ \frac{{\rm Joule}}{{\rm meter}^3} .
\eeq
If we assume $ \gcal \approx 1$,    the mass of the  lightest particle is $m_1 = 0.003 \, {\rm eV}$.   It is worth noting that this is in the ballpark  
of proposed  neutrino masses\footnote{The literature on neutrino masses is large and mainly unfamiliar to us;  we can however cite the review \cite{neutrino} and references therein.}.   It has certainly been noticed before that $\rhovac \sim  m^4$ with $m$ the neutrino mass,   but this was previously just a curious observation,  without derivation,    and perhaps just a coincidence.       We  suggest that our analogous results above in $2d$,   which are exact for integrable theories,     can possibly explain this $4d$ observation.

\bigskip

\noindent  {\bf Note added.} ~~ Upon completion of this  work we were informed\,\footnote{We thank M. Montero and G. Venken for private communications.}   of interesting recent work that provides some indirect evidence supporting  our proposal 
in $4d$,  namely equation \eqref{rhovac4d}.      From a very different approach involving charged black holes and the notion of a Swampland 
 \cite{Montero1,Montero2},   it was proposed that 
\beq
\label{Montero}
\rhovac <    \frac{m^4}{2 e^2}
\eeq
where $m$ is the mass of a charged particle,  and $\alpha = e^2/4 \pi  $  is the electromagnetic  fine structure constant.      This is weaker than \eqref{rhovac4d} since it is an upper bound rather than an equality.    However quite   remarkably it is consistent with \eqref{rhovac4d} if $m$ in \eqref{Montero} is the lightest mass particle and $<$ is replaced with $\leq$. 
In other words the novelty of our proposal \eqref{rhovac4d} is that whereas it is consistent with \eqref{Montero} if $m_1$ is the lightest mass,  it proposes that the lightest mass particle saturates the inequality leading to an equality.  
This result provides some optimism that our proposal can be derived directly in the 4d quantum field theory.   It would thus also be interesting to derive our exact 2d results from this
blackhole/swampland approach.

\section*{Acknowledgments}

We  wish to thank Changrim Ahn  for discussions,   as our work together \cite{Changrim} indirectly  led  in part to the considerations of this paper.   
We also wish to thank Miguel Montero and  Gerben Venken for pointing out their  potentially related work  \cite{Montero1,Montero2} after this work was originally completed.   
Finally we thank the referee at JHEP for several expert and insightful remarks that substantially improved the original version of this article.

\vfill\eject

\end{document}